\documentstyle[12pt]{article}
\def\be{\begin{equation}}
\def\bea{\begin{eqnarray}}

\def\ee{\end{equation}}
\def\eea{\end{eqnarray}}
\def\R{\rm {I\kern-.200em R}}
\def\C{\rm {I\kern-.520em C}}
\begin{document}
\begin{titlepage}
\begin {center}
{\large { 2-d Gravity as a Limit of the \\SL(2,R) Black Hole }}
\\
\vskip 1cm{M. Alimohammadi $^{(a,b)}$ and F.Ardalan $^{(a,c,d)}$ }\\
\vskip 2cm
 {\it $^a$ Institute for Studies in Theoretical Physics and
Mathematics\\  P.O.Box 19395-5746, Tehran, Iran}\\
{\it $^b$ Physics Department, Tehran University , North Karegar ,\\
 Tehran, Iran}\\
$^c$ {\it Physics Department, Sharif University of Technology} \\
{\it P.O.Box 11365-9161, Tehran, Iran}\\
{\it $^d$ Department of Physics, University of Cincinnati,\\
 Cincinnati, Ohio 45221, U.S.A.}\\
\end {center}
\vskip 1cm
\begin{abstract}
The transformation  of the $SL(2,R)/U(1)$ black hole under a boost of the
subgroup $U(1)$ is studied. It is found that the tachyon vertex operators
of the black hole go into those of the $c=1$ conformal field theory coupled
to gravity. The discrete states of the black hole also tend to the discrete
states of the 2-d gravity theory. The fate of the extra discrete states of
the black hole under boost are discussed.
\end{abstract}
\vskip 2cm
\vfill
\end{titlepage}
\noindent
\section{Introduction}
The relation between the two simple string theory models in two dimensions,
the critical $U(1)$ gauged WZW $SL(2,R)$ model $^{[1-4]}$, and the noncritical
string theory of a one dimensional matter field coupled to Liouville field,
has attracted considerable attention $^{[5-14]}$.

   In Ref.[1] it was argued that as it is not possible to remove one of the
parameters  of the two dimensional black hole in favour of the Liouville field
in all the regions of the black hole geometry, the theory can not be regarded
as a  non-critical string theory of $ c=1$ matter coupled to gravity. In
agreement with this result, Distler and Nelson$^{[6]}$ studied the BRST
cohomology of the black hole and found that there are more discrete states in
the black hole
than in the Liouville theory. Also by looking at the behaviour of the states
of the black hole near the horizon, it was found $^{[12]}$ that there are only
a
few states that do not diverge near the horizon and are therefore  physical,
the $W_\infty$ states  not being among them.

   On the other hand using the free field realization of ${ SL(2,R)/ U(1)}$
and the true BRST charge in the black hole, it was shown that as far as the
energy-momentum tensor is concerened, the model is identical to 2d gravity
$^{[8]}$. It was then claimed that the extra states that appear in the
$SL(2,R)/U(1)$ , are BRST exact and therefore the spectrum of both theories
are  the same. It has been argued that $^{[13]}$ there are null
states in the black hole which lead to even more discrete states than in
Ref.[6]. Yet
in a different approach the number of discrete states of the black hole come
out even less than those of 2-d gravity $^{[12]}$. On the other hand in the
context of matrix models, it has been shown that the 2 dimensional black hole
theory and the 2 dimensional gravity theory are closely related $^{[14]}$.
Therefore the relation between the two theories warrants further investigation.

   In this work  we will study the relation by a different method, i.e., by
gauging the SL(2,R) by a nilpotent subgroup and looking at the behaviour of the
black hole theory ${SL2,R)/U(1)}$ when the ${U(1)}$ tends towards this
nilpotent
subgroup, which we call ${E(1)}$.
   When the ${U(1)}$ subgroup of the ${SL(2,R)/U(1)}$ black hole is substituted
with the subgroup ${E(1)}$, it is found $^{[15,16]}$ that the resulting theory
is nothing but the one dimensional Liouville field theory with zero
cosmological
constant $^{[18]}$. The same result can of course be obtained by boosting
the original ${U(1)}$ subgroup and letting the boost parameter go to infinity.
This reduction of the degrees of freedom from two to one, has been studied and
is understood to be a consequence of an enlarged symmetry $^{[17]}$.
   In the effort to understand the details of the effect of the boost on the
black hole theory, the effective Lagrangian of the
boosted theory was studied in the limit of large but finite boost parameters,
and found to resemble the corresponding quantity in the theory of a c=1
conformal matter coupled to Liouville field$^{[15,16]}$.

   In this paper we have pursued this line of investigation in more detail
and have found that not only the action and the tachyon of the black hole
tend to those of 2-d gravity, but also the discrete states of the black hole
tend to the discrete states of the 2-d gravity. The identification of the
quantum numbers of the former theory with the momenta of the latter$^{[6]}$,
will then appear as a natural consequence of the boost transformation. The
transformation also explains the disappearance of the extra discrete states
which occur in black hole and not in the 2-d gravity.

    In section 2 we will review the results of Ref.[16] for E(1) gauging and
discuss the free field representation of the nilpotent gauged WZW model of
$SL(2,R)$ and show that the stress tensor of this model is the same as in the
Liouville theory. In section 3 we will study the limit of the primary fields of
$SL(2,R)/U_t(1)$ at $t \rightarrow \infty$, where t is the boost parameter, and
show that the primary fields in the regions V and III of Ref.[1] lead to the
vertex operators of the c=1 theory coupled to Liouville. In section 4 we study
the operator aspects of the the boost transformation and in section 5 we will
take up the discrete states.

\section {Gauging SL(2,R) by its Nilpotent Subgroup}

Let us take the following parametrization for elements $g \in SL(2,R)$ ,
 \be g=  \left( \begin{array}{ll}  a&u\\-v&b  \end{array} \right),\ \ \ \ \
\ \ \ ab+uv=1. \ee
We consider the nilpotent subgroup E(1) of SL(2,R) generated by $\sigma =
\sigma_3+i \sigma_2$ and use the axial gauge freedom ( $ g \rightarrow
hgh$ ) to fix the gauge by the following condition :
\be a+b=0, \ee
which is valid in region V . Then using the parameters:
$$x={1 \over 2} (u-v)$$
\be e^{\varphi '} = {1 \over 4} (u+v-2a), \ee
we find the following effective action $^{[16]}$
\be I_{eff} = \frac {k}{4 \pi} \int d^2 \sigma \sqrt {h} h^{ij} \partial_i
\varphi' \partial_j \varphi'  +\frac {1}{4 \pi}\int d^2 \sigma \sqrt {h}
R^{(2)}\varphi' . \ee
In the above equation $h^{ij}$ is the two dimensional metric
and $R^{(2)}$ is the curvature of the world-sheet .
This one-dimensional action is nothing but the Liouville action . We will now
show this equivalence of $SL(2,R) / E(1) $
and Liouville theory
at the level of stress tensor . As is well known,
if one uses the Gauss
decomposition to represent the group elements of $SL(2,R)$, the following
representations for the currents of $SL(2,R)_k$ in terms of free fields $\beta$
,$\gamma$ and $\phi $ can be obtained $^{[19]}$:
\pagebreak
$$ J_+ = \beta $$
\be J_- = \beta \gamma^2 +\sqrt {2k'} \gamma \partial \phi +k \partial
\gamma \ \ \ ,  \ \ \ \ k'=k-2 \ee
$$J_3= -\beta \gamma - \sqrt {k'\over 2} \partial \phi $$
where $\beta$ and $\gamma$ are the commuting ghost fields with dimensions
$h=1,0$ and with OPE's $\beta (z) \gamma (w) \sim {1\over z-w}$ and $\phi (z)
\phi (w) \sim -\lg (z-w)$. Then using Sugawara construction,
the stress tensor of $SL(2,R)_k$ becomes:
\be T _{SL(2,R)}(z) = \beta \partial \gamma -{1\over 2} (\partial \phi)^2
-{1\over \sqrt {2k'}} \partial ^2 \phi \ee
Now we want to gauge away the nilpotent subgroup of $SL(2,R)$, i.e. $J_+$, by
using the BRST method. As $J_+ (z) J_+ (w) $ is regular, there is no need to
introduce a gauge field
(auxiliary field) for constructing the BRST charge ($Q _+$), and hence there is
no need to introduce ghosts to fix the gauge field. In this way we arrive at
the following
expression for the BRST charge of the nilpotent subgroup of $SL(2,R)$:
\be Q_+ = \oint dz J_+ (z) \ee
which satisfies,
$$Q_+ ^2=0$$.
As we do not introduce the gauge field, so it has no contribution in $T(z)$
and therefore
 \be T_{SL(2,R)/ E(1)}=\beta \partial \gamma -{1\over 2} (\partial
\phi)^2 -{1\over \sqrt {2k'}} \partial ^2\phi \ee
But there are terms in Eq.(8) which are BRST exact and must be subtracted
from the stress tensor. It can be easily shown that:
$$\beta \partial \gamma =\partial (\gamma \beta) - \gamma \partial \beta$$
$$= \partial (\gamma \beta) -\{Q_+, {1\over 2} \partial \beta \gamma^2\}$$
 $$= \{Q_+, \partial ({1\over 2} \beta \gamma^2)\} - \{Q_+, {1\over 2} \partial
\beta \gamma^2\}$$
Thus up to a BRST exact term and at the level $k={9\over 4}$, we find that:
\be T_{SL(2,R)/E(1)} = -{1\over 2}(\partial \phi)^2 -\sqrt {2}
\partial ^2 \phi \ee
The above equation is exactly the Liouville action at zero cosmological
constant.The same result can be obtained if we consider the stress tensor of
$SL(2,R) / U_t(1) $ and look at its behaviour at $t \rightarrow \infty $ .
\section {Primary Fields}
The primary fields of WZW model are defined via its matrix elements. In the
case
of $SL(2,R)$ gauged by $\sigma_3$, the vertex operator in the region I is
$^{[3]}$:
\be V_\lambda ^\omega = <\lambda ,\omega|g(y,\tau)|\lambda, -\omega> \ee
where $ \lambda$ defines the spin of  $SL(2,R)$  representation, $\omega$ is
the eigenvalue of $\sigma_3$ and $y$ and $\tau$ are defined by:
\be y=uv  ,\ \ \ \ \ \ \ \    e^{2\tau}= -{u \over v}. \ee
In this region the suitable gauge condition is  $a-b=0$ .
   There are four different vertex operators in the region I which have
different behaviours near the horizon and infinity, among which one, denoted
$U_\lambda ^\omega$, can be naturally extended to the region III,$^{[3]}$ :
\be  U_\lambda ^\omega= e^{-2i\omega\tau} F_\omega ^ \lambda (y)\\=
e^{-2i\omega\tau} (-y)^{-i\omega}
B(\nu_+, \bar\nu_-),F(\nu _ +, \bar\nu_-,1-2 i\omega,y) \ee
where $B(\alpha, \beta )=\Gamma ( \alpha ) \Gamma ( \beta ) /
\Gamma(\alpha + \beta )$, F is the hypergeometric function $_2F_1$ , and
\be \nu_\pm = {1\over 2} -i(\lambda \pm \omega) \ee
   However, it is convenient to work with the boosted group element
$g$ , in Eq.(10), rather than using the state corresponding to the $U_t(1)$. We
therefore have,
\be U_\lambda ^\omega (t)=<\lambda,\omega|g(y_{-t}, \tau_{-t})|\lambda ,
-\omega>   \ee
where
\be g_{-t}=e^{{t\over 2} \sigma _1}g e^{-{t \over 2 }\sigma _1}  \ee
   If the eigenvalues of $\sigma=\sigma_3+i\sigma _2$ are denoted by $\chi$
$$\sigma|\lambda ,\chi >=\chi |\lambda ,\chi>$$
then as $\sigma^t_3 \stackrel {t \rightarrow \infty} \longrightarrow e^t
\sigma$ the states $|\lambda ,\omega>_t$ must tend
to $|\lambda ,\chi>$ as $t\rightarrow \infty$. As a result,
\be \omega=\chi e^t \ee
and
\be \nu_ \pm= {1\over 2}-i(\lambda \pm \chi e^t) \ee
   To find the limit of the vertex operators , we will express the
hypergeometric function in terms of the associated Legendre function  of the
$2^{nd}$ kind $Q_\mu ^\nu(z) $ $ ^{[20]}$, and obtain,
$$U_\lambda ^\chi (t)\stackrel {t\rightarrow \infty}\longrightarrow e^{-4i\chi
{u+v \over u-v}}Q_ \nu ^0(1+{8\over(u-v)^2})$$ which when
$|\nu| \rightarrow \infty$, reduces to:
 \be U_\lambda ^\chi (t)\rightarrow e^{-4i\chi {u+v\over u-v}}
\sqrt {\pi \over 2}(w^2-1)^{-1\over 4}(w-\sqrt
{w^2-1})^{-{i\over 2}\chi e^t} \ee
where $w=1+{8\over (u-v)^2}$.
As expected, there is no connection to 2-d gravity in the region I.

The vertex operators in the region V are $^{[3]}$ :
\be W_\omega ^\lambda (y,\tau)=e^{-2i\omega\tau} y^{-i\omega} F(\nu_+ , \bar
\nu_- , 1 , 1-y) \ee
where $y=uv$ and $\tau= {1\over 2}\ln ({u/v})$ and the gauge condition is
$a+b=0$.
As this function has no singularity when crossing the singularity at $y=1$, it
can be trivially continued to region III.
In the same way as discussed in the previous section, the corresponding vertex
operator
of ${SL(2,R)/ U_t(1)}$ in this region can be recovered from Eq.(19)
by simply transforming $y \rightarrow y_ {-t}$
, $\tau \rightarrow \tau _{-t}$ and taking the gauge condition
$a_{-t}+b_{-t} =0$
.After some algbra, using a similar set of identities as in the case of region
I, we obtain,
\be W_\chi^\lambda(t) \stackrel {t\rightarrow \infty} \longrightarrow
 {1\over \pi}e^{-2ix\chi e^{-\varphi'}}e^{\pi \chi e^t}
e^{-(t+\varphi')}K_{2i\lambda}(2\chi e^{-\varphi'})\ee
Fortunately the dependence of $W_\chi ^\lambda$ on $t$ is such that we can
absorb it consistently in $\varphi'$, and
therefore if we define $\varphi=\varphi' +t$ and $xe^{-\varphi}=X$  and use
Eq.(16), we will finally obtain,
\be W_\omega^\lambda (t) \stackrel {t\rightarrow \infty}\longrightarrow
{e^{\pi \omega}\over \pi}e^{-2i\omega X}
e^{-\varphi}K_{2i\lambda}(2 \omega e^{-\varphi}) \ee
The Eq.(21) is exactly the vertex operator of c=1 coupled to 2-d gravity with
non-zero
cosmological constant $^{[21]}$. This equivalence is even clearer when the
vertex operator
is considered on-shell, that is when $\lambda =\pm {\omega / 3}$ $^{[3]}$.
Eq.(20) also shows that the eigenvalue of $\sigma$
plays the role of the cosmological constant  $\chi =\sqrt {\mu}$
$^{[3]}$.
   When $\chi \rightarrow 0$, we obtain,
\be W_\omega^\lambda (t) \stackrel {t\rightarrow \infty} \longrightarrow
e^{-(\varphi'+t)} e^{-ix\chi e^{-\varphi'}} \{Ae^{2i\lambda(\varphi' +t)}
+c.c.\} \ee
where:
$$A={\Gamma (2i\lambda) \over \Gamma ({1\over2}+i\lambda +i\omega)\Gamma
({1\over 2}+i\lambda - i\omega)}$$
As in the previous case, if we define $\varphi =\varphi' + t$ and
$X =xe^{-\varphi}$
we will arrive at the following expression for the primary fields
\be W _\omega^\lambda \rightarrow e^{-\varphi} e^{-2i\omega X } \{
Ae^{2i\lambda \varphi}+A^* e^{-2i\lambda \varphi}\} \ee
However, this is nothing but the vertex operator of c=1 plus Liouville at zero
cosmological
constant, of course after applying the on-shell condition. The
scattering matrix is ${A/ A^*}$.
Therefore $\chi $ plays exactly the role of the cosmological constant
and the boosted black hole in region V is equivalent to 2-d gravity.
At $t=\infty$ where $\chi =\omega e^{-t}=0$, Eq.(22) becomes:
 \be W_\omega^\lambda (t=\infty) = Ae^{2i(\lambda -1)\varphi} +c.c. \ee
which is the expression for
the vertex operator of pure gravity, Liouville theory, as expected:
as $t \rightarrow \infty$, $\sigma _3^t \rightarrow \sigma$ and
we expect the theory to reduce
to $SL(2,R) / E(1) $, which is the Liouville theory.
\section { Limit of the Operators}
In the region V where the gauge condition is $a+b=0$ ,the effective action is
$^{[1]}$ :
\be I_{eff}=-{k \over 4\pi}\int {\partial u  \bar\partial v +
\partial v  \bar\partial u \over 1-uv}d^2z+{1\over 4\pi}\int d^2\sigma
\sqrt {h}R^{(2)} ln(1-uv). \ee
If we boost this action and keep the next to leading order terms in $\epsilon=
e^{-t}$, we will find :
$$ I_{eff} = \frac {k}{4 \pi} \int d^2 \sigma \sqrt {h} h^{ij} [
\partial_i \varphi \partial_j \varphi -(\partial_i X \partial_j X +
X \partial_i \varphi \partial_j X)]$$
\be +\frac {1}{2 \pi}\int d^2 \sigma \sqrt {h}
R^{(2)}(\varphi - {1 \over 4}X^2), \ee
where $X=xe^{-\varphi}$ and $\varphi=\varphi'+t$, which resembles the theory of
a matter field X coupled to a Liouville field ${\varphi}$, including an
interaction term $^{[5]}$; which we may ignore if we assume that X is small,
because of its ${\epsilon}$ dependence, and if ${\partial X}$ is comparable to
${\partial \varphi}$.
   Note that if we redefine the Liouville field as,
\be \Phi=\varphi - {1 \over 4} X^2 ,\ee
we obtain,
\be I_{eff} = \frac {k}{4 \pi} \int d^2 \sigma \sqrt {h} h^{ij} (
\partial_i \Phi \partial_j \Phi -\partial_i X \partial_j X)
 +\frac {1}{2 \pi}\int d^2 \sigma \sqrt {h}
R^{(2)}\Phi, \ee
which indeed is the standard 2-d gravity action. Observe that the correction to
${\varphi}$ is only second order in ${\epsilon}$ and will not affect the
results for the tachyon vertex operators of previous section.
To go further we need to know the operator properties of the fields in the
boosted theory.

 To investigate the OPE's of these fields, it is necessary to consider the free
field representation of SL(2,R). As it is known, if we use the Gauss
decomposition to parametrize the SL(2,R) group elements , that is :
\be g=e^{\gamma \sigma_+}e^{\phi'\sigma_3}e^{\chi \sigma_-}=
\left( \begin{array}{ll}  e^{\phi'}+\chi\gamma e^{-\phi'}&\gamma e^{-\phi'}\\
\chi e^{-\phi'} & e^{-\phi'} \end{array} \right) \ee
where $\sigma_\pm={1 \over 2}(\sigma_1\pm i \sigma_2)$ , then the currents
$J=J_i\sigma_i=-k(\partial g)g^{-1}$ reduce to the representation (5) with
definition of $\phi$ and $\beta$ as :
$$\phi=-k\sqrt{2\over k'}\phi'$$
\be \beta=-k\partial \chi e^{-2\phi'} . \ee
Boosting the representation (29) and imposing the gauge condition, results in
the following relations for the leading terms of $\beta_{-t},\gamma_{
-t}$ and $\phi_{-t}$ :
$$\beta_{-t}=k(x\partial \varphi - \partial x)e^\varphi=-k\partial X e^{2
\varphi}$$
\be \gamma_{-t}=1+xe^{-\varphi}=1+X \ee
$$\phi_{-t}=k\sqrt{2\over k'}\varphi$$
  Now the OPE's of the above fields are :
$$\beta_{-t}(z) \gamma_{-t}(w) \sim {1\over z-w} \ \ \ \ \ \ , \ \ \ \ \
\phi_{-t}(z) \phi_{-t}(w) \sim -lg(z-w) $$
\be \beta_{-t}(z) \phi_{-t}(w) \sim \gamma_{-t}(z) \phi_{-t}(w) \sim regular\ee
therefore using the Eq.(31), the following OPE's are dictated for $X$ and
$\varphi$
fields :
$$\varphi(z)\varphi(w) \sim -{k'\over 2k^2}lg(z-w)$$
\be \varphi(z) X(w) \sim regular \ee
$$<X(z)X(w)>=-{1\over 2k}lg(z-w).$$
The above equations show that the interpretation of $\varphi$ and $X$ as the
Liouville and the c=1 bosonic fields, which commute with each other, is
permitted.

\section {Discrete States}

   The discrete states of the $SL(2,R)/U(1)$ black hole were found in reference
[6] using the parafermionic modules $V_{j,m}$ built on the states $U_{j,m}$ of
the discrete irreducible representations of $SL(2,R)$, where $j$ and $m$ are
the
usual angular momentum lables of $SL(2,R)$. It was found that aside from the
propagating tachyon states with $$m=\pm{3(j+1)/2},$$ there are three sets of
discrete states labled as $D$ ,$ C$ and $\hat{D}$ with
$$m=\pm {3\over 8}(2s-4r-1) \ \ \ \ , \ \ \ \     j={1 \over 8}(2s+4r-5)$$
for $\hat{D}$ states; and
$$m=\pm{3\over 4}(s-2r+1) \ \ \ \ , \ \ \ \ \      j={1\over 4}(s+2r-3)$$ for
$D$ states; and
$$m={3\over 2}(s-r) \ \ \ \ , \ \ \ \      j={1\over 2}(s+r-1)$$ for the $C$
states,
where $s$ and $r$ are positive integers.
{}From the form of the discrete states of the 2-d gravity theory,
$$p_x=\sqrt{2}(p-q) \ \ \ \ ,  \ \ \ \ p_\varphi=\sqrt{2}(p+q-1),$$
 with $p$ and $q$ positive integers ,
it was suggested that the sets $D$ and $C$ correspond to the 2-d gravity
discrete
 states with the identification
\be p_x=2\sqrt{2}m/3 \ \ \ \ , \ \ \ \ \    p_\varphi=2\sqrt{2}j;\ee
and that the $\hat{D}$ did not correspond to any 2-d gravity states.
   These extra states will of course not appear if a free field representation
of $SL(2,R)/U(1)$ is used, as the operator structure and the energy momentum
tensor of the theories are not distinguishable in this representation. However,
once the Kac-Moody Verma modules are usd, the extra states are inescapable.
Below, we will see that performing a boost transformation on the states to
the black hole will remove the $\hat{D}$ states and reduce the spectrum to that
of the 2-d gravity.

   To begin with, recall that the states $\hat{D}$ and $D$ are related by the
screening operators,
$$S^{\pm}=\oint dze^{\sqrt{k'/2}\phi^1 \pm i \sqrt{k/2}\phi^2} ,$$
in terms of the free fields $\phi^{1}$ and $\phi^{2}$. In fact
$\hat{D}$ = ker $(S)$ and $D=V/\hat{D}$, and the modules $V$ are generated on
the
SL(2,R) base representation operators,
$$U_{j,m}=e^{j\sqrt{2/k'}\phi^{1}+m\sqrt{2/k}(i\phi^{2}+\phi^{3})},$$
where $\phi^{3}$ is the $U(1)$ free field.
   We now apply our boosting map on these operators and obtain,
$$U_{j,m} \rightarrow const.  e^{-2j\frac{k}{k'}\varphi' - 2(j+m)X},$$
$$S^+ \rightarrow const. \oint dze^{-k\varphi'-2(k-1)X},$$
\be S^- \rightarrow const. \oint dze^{-k\varphi' -2x}.\ee
{}From these equations, as we have interpreted $\varphi$ to be the
Liouville field, considering the normalisation Eq.(33) of $ \varphi$, we see
that
\be p_{\varphi}=2 \sqrt {2}j,\ee
as suggested in Ref.[6]. The corresponding relation for $p_x$,
\be p_{x}= 2\sqrt {2}(m+j)/3 ,\ee
has the coefficient suggested in Ref.[6], and $m$ substituted for $m+j$.

   Next, taking the OPE's of the screening operators with the Verma module
operators, we see that in the approximation we are considering, the screening
operators
fail to be well defined, indicating that the states $\hat{D}$ are removed from
the spectrum.
   To confirm the result above we have calculated the commutator of the limit
of the screening operators with that of the Virasoro operator $L_{0}$, and
they also fail to vanish in the first order approximation theory.
   We conclude that in the large boost limit of the black hole theory, the
extra discrete states disappear and the spectrum becomes that of the 2-d
gravity.

{\bf Acknowledgements}
We would like to thank A.Morozov for valuable discussions on the free field
representation of GWZW models.
\begin{center}
{\large  References \\}
\end{center}



\begin{enumerate}
\bibitem{}  E. Witten, Phys. Rev. D44(1991),314.
\bibitem{}  G. Mandel, A. M. Sengupta and S. R. Wadia, Mod. Phys. Lett.
A6(1991)
,168.
\bibitem{}  R. Dijkgraaf, H. Verlinde and E. Verlinde, Nucl. Phys.B 371(1992)
 269.
\bibitem{}  A. A. Tseytlin, Nucl. Phys. B399(1993)601; I. Bars, Phys. Lett.
B293(1992)315.
\bibitem{}  E. J. Martinec and S. L. Shatashvili, Nucl. Phys. B368(1992)338;\\
 M.Bershadsky and D. Kutasov, Phys. Lett. B266(1991),345.
\bibitem{}  J. Distler and P. Nelson, Nucl. Phys. B374(1992),123.
\bibitem{}  S. Mukhi and C. Vafa, Nucl. Phys. B407(1993),667.
\bibitem{}  T. Eguchi, H. Kanno and S. K. Yang, Phys. Lett. B298(1993),73.
\bibitem{}  T. Eguchi, Phys. Lett. B316(1993),74.
\bibitem{}  M. Ishikawa and M. Kato, Phys. Lett. B302(1993),209.
\bibitem{}  S. Chaudhuri and J. D. Lykken, Nucl. Phys. B396(1993),270 \\ K.
Becker
            and M. Becker : Interactions in the $SL(2,R)/U(1)$ blackhole
            background , CERN-TH.6976/93.
\bibitem{}  N. Marcus and Y. Oz, hepth/9305003.
\bibitem{}  K. Itoh, H. Kunitomo, N. Ohta and M. Sakaguchi; BRST Analysis of
Physical States in Two-Dimensional Blackhole, OS-GE 28-93.
\bibitem{}  S. R. Wadia, hepth/9503125
\bibitem{}  F. Ardalan, ``2D black holes and 2D gravity'', in Low Dimensional
Topology and Quantum Field Theory, Ed.,H. Osborn, (Plenum Press,1993), p.177
\bibitem{}  M. Alimohammadi, F. Ardalan and H. Arfaei, Int. J. Mod.
Phys. A10(1995)115.
\bibitem{} F. Ardalan and M. Ghezelbash, Mod. Phys. Lett. A9(1994)3749.
\bibitem{}  P. Forgacs, A. Wipf, J. Balog, L. Feher and L. O'Raifeartaigh,
Phys.
 Lett. B227(1989),214 \\
 L. O'Raifeartaigh, P. Ruelle and I. Tsutsui, Phys. Lett. B258(1991),359 \\
 A. Alekseev and S. Shatashvili, Nucl. Phys. B323(1989),719 \\
 M. Bershadsky and H. Ooguri, Commun. Math. Phys. 126(1989),49. \\
 G. T. Horowitz and A. A. Tseytlin, Phys. Rev. D50(1994),5204. \\
 C. Klimcik, Null gauged WZNW theories and Toda-like $\sigma$-models,
 hep-th/9501091.
\bibitem{}  A. Gerasimov, A. Morozov and M. Olshanetsky, Int. J. of Mod.
Phys.A5
         (1990),2495.
\bibitem{}  Erdely, Magnus, Oberhettinger and Tricomi: Higher Transcendental
Functions (Bateman Manuscript Project), McGraw-Hill (1953).

\bibitem{}  G. Moore, N. Seiberg and M. Staudacher: From Loops to States in 2D
       Quantum Gravity, Rutgers preprint RU-91-11 (March,1991).

\end{enumerate}
\end{document}